\documentclass{article}

\usepackage{amssymb}
\usepackage{enumerate}
\usepackage{a4wide}
\usepackage{graphicx}
\usepackage{clrscode}
\usepackage{float}
\usepackage{epsfig}
\usepackage{subfigure}
\usepackage{path}

\newcommand\ignore[1]{}
\newcommand{\dm}{dm}

\newcounter{algo}
\renewcommand{\thealgo}{\arabic{algo}}

\newlength{\defbaselineskip}
\newcommand{\setlinespacing}[1]%
           {\setlength{\baselineskip}{#1 \defbaselineskip}}

\newcommand{\cgal}{{\sc Cgal}}

\newcommand{\TSmall}{{\sc Small}}
\newcommand{\TMedium}{{\sc Medium}}
\newcommand{\TLarge}{{\sc Large}}
\newcommand{\CC}{C\raise.08ex\hbox{\tt ++}}
\newcommand{\ft}{\;\rm ft}

\begin{document}

\title{Fully Automatic Trunk Packing with Free Placements}
\author{Ernst Althaus\thanks{Johannes Gutenberg Universit\"at Mainz, Staudingerweg 9, 55099 Mainz, Germany
{\tt ernst.althaus@uni-mainz.de}} \and Peter
Hachenberger\thanks{MADALGO (Massive Data Algorithmics, a Center of the Danish National Research
Aarhus, Denmark, {\tt phachenb@madalgo.au.dk}}
\thanks{PH was supported by the
Netherlands' Organisation for Scientific Research (NWO) under project
no.~639.023.301.}}


\maketitle

\begin{abstract}
We present a new algorithm to compute the volume of a trunk according
to the SAE J1100 standard. Our new algorithm uses state-of-the-art
methods from computational geometry and from combinatorial
optimization. It finds better solutions than previous approaches for
small trunks.
\end{abstract}

\section{Introduction}
\label{sec:introduction}

The volume of the trunk of a car has a significant influence on the
car design process. In Germany the volume is measured according to the
DIN 70020 standard and in the USA according to the SAE J1100 standard.
The DIN 70020 standard asks to pack as many 1-liter-boxes of size
$20cm\times 10cm\times 5cm$ as possible into the trunk. The total
volume of the packed boxes then defines the volume of the trunk.

\begin{figure}[t]
\center
\includegraphics[width=0.53\textwidth]{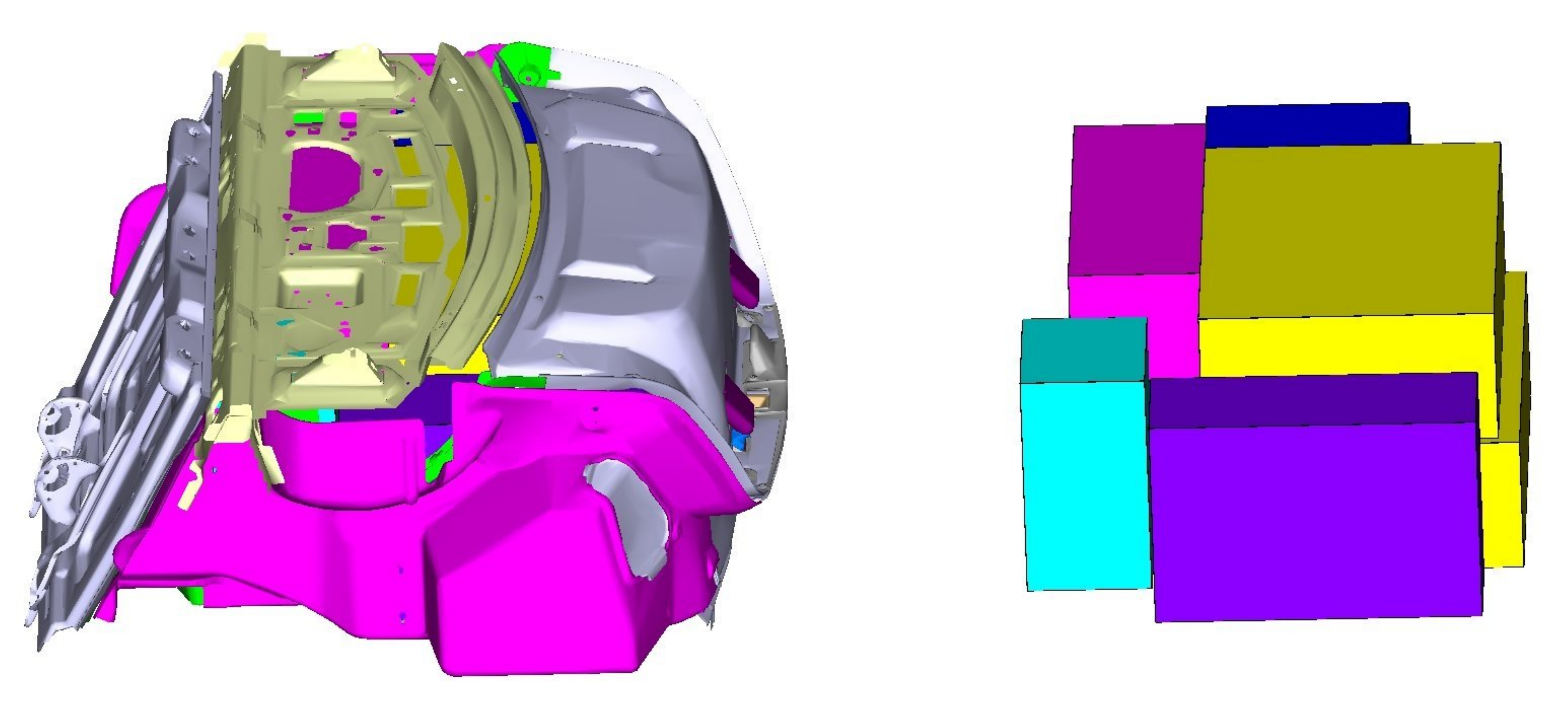}
\hfill
\begin{tabular}[b]{|c|r|c|}
\hline
ID & \multicolumn{1}{|l|}{\ Size} & max \\
\hline
A & 229mm $\times$ 483mm $\times$ 610mm & 4 \\
B & 165mm $\times$ 330mm $\times$ 457mm & 4 \\
C & 229mm $\times$ 406mm $\times$ 660mm & 2 \\
D & 216mm $\times$ 457mm $\times$ 533mm & 2 \\
E & 203mm $\times$ 229mm $\times$ 381mm & 2 \\
F & 178mm $\times$ 356mm $\times$ 533mm & 2 \\
\hline
G & 1143mm $\times$ 204mm $\times$ 204mm & 2 \\
\hline
H & 152mm $\times$ 114mm $\times$ 325mm & 20 \\
\hline
\end{tabular}
\caption{
\label{fig:boxset}
According to the SAE J1100 standard, a packing consists of suitcases
taken from a fixed set. There are six different types of suitcases
(A--F), a golf bag (G), and the H-box, which represents additional
loose baggage. Some suitcases are allowed twice in the packing, others
four times. There may be up to 20 H-boxes, but only in addition to an
optimal packing of boxes A--G. Left: Trunk. Middle: Packing. Right:
Predefined types of boxes.}
\end{figure}

In this paper we focus on the SAE J1100 standard. It asks to pack
cuboid suitcases, from now on also denoted as boxes, and measures the
total volume of the suitcases. The boxes are taken from a fixed set of
six types of suitcases, a golf bag, and the so-called H-box, which
represents loose baggage. Each box type can only be packed a limited
number of times---usually two or four times. The packing is performed
in two steps. In the first step it is not allowed to pack any
H-box. Only after obtaining an packing with respect to the other
boxes, up to 20 H-boxes may be added in a second
step. Figure~\ref{fig:boxset} shows a trunk together with a packing
and summarizes the important properties.

Until a few years ago, evaluating the size of a trunk was only done
manually. Especially for the DIN 70020 standard this is a very tedious
task that takes much time and manpower. In order to reduce the cost
and to be able to determine the volume of a trunk early in the design
process, car manufactures are interested in automated methods. A
German car manufacturer sparked the recent interest in such methods,
first for the DIN 70020 standard
(\cite{efkrs-05,Reichel2003,Karr04,Neum06,Reichel06,Ries05}), and
later also for the SAE J1100 standard (\cite{absw-07,cd-03}). The
manufacturer asked for an automatic method that provides a solution
not worse than 98\% of the best manually obtained packing within 24 hours.

Almost all methods rely on discretizing the trunk, i.e., aligning the
trunk with a uniform grid and finding a packing by aligning the boxes
with the grid. For the DIN 70020 standard the grid method is the most
reasonable approach, because the configuration space, i.e., the set of
potential packings, is too large to be tested, if free placement is
allowed. The bigger boxes of the SAE J1100 standard yields
considerably smaller configuration spaces than the DIN 70020 standard.
Still, it is necessary to at least discretize the orientation of the
boxes. There has been no research on modeling configuration spaces for
objects in three-dimensional space with free placement and
orientation, yet. Even the two-dimensional counterpart has been
investigated only briefly. The problem is inherently complex.

The major problem of the grid approach for the SAE J1100 standard is
to find a suitable grid. The measures of the boxes have non-simple
ratios. To make each box fit exactly into the grid, the grid cells
must be very small, which increases the complexity of the optimization
problem, and therefore also the running time. The grid approach does
not pay off any more~\cite{bsw,Reichel06}. Althaus~\emph{et al.}
handle the problem by working with a reasonable grid size and allowing
the boxes to overlap slightly. The overlap is then resolved with a
physical simulation of the contacts~\cite{bsw}. Their algorithm is not
guaranteed to find a packing in each run. Instead, they perform
several runs, each of which transforms a set of overlapping boxes into
a packing. If at least one packing is found, the best packing is
returned.

In the following we give an algorithm that does not rely on a grid.
We discretize the orientation of the suitcases in the same way as in
the grid approach, i.e., we restrict the orientations to the six
axis-aligned directions, but our approach allows free positioning of
the suitcases within the trunk. Without the grid, configuration space
and its description grows considerably. On the other hand, our
approach allows to reliably find good packings. Moreover, our approach
consists of several tasks, many of which can be performed in
parallel. The computation time can be reduced considerably by using
multiple machines.

\subsection{Outline of Our Algorithm}
\label{sec:algorithm}

Reichel proved that the trunk packing problem is
NP-hard~\cite{Reichel06}. Thus, we cannot hope for finding the optimal
solution. Our algorithm is based on enumeration. As there are
infinitely many different packings, our algorithm does not enumerate
packings, but so called packing patterns as introduced by Schepers for
the problem of packing small boxes into a large
box~\cite{schepers}. Such a pattern bundles a class of packings by
fixing a small set of properties. In our scenario, enumerating packing
patterns has two advantages:
\begin{enumerate}
\item There are only a finite number of packing patterns. Thus, the
search tree becomes finite.
\item The properties fixed by a pattern limit the maximum volume of a feasible packing.
\end{enumerate}
\noindent Knowing the maximum feasible volume of a packing pattern
allows us to prune the search when the maximum feasible volume is not
large enough to be interesting. If a pattern is interesting it
suffices to check whether there is any feasible packing for the
pattern. We solve the feasibility tests by solving a linear program.

The biggest challenge of our approach is to determine a linear program
of manageable size. We start by characterizing the feasible positions
of the suitcases. For a fixed orientation the area of all feasible
positions of suitcase $F$ can be computed with the Minkowski sum, as
will be explained in Section~\ref{sec:feasiblearea}. Computing the
Minkowski sum for each of the six orientations we obtain six
polyhedra, which together represent the feasible configurations of
$F$.

In a linear program it is preferable to work with convex regions,
which can be described as a set of linear inequalities. Then a point
is inside the region iff all inequalities are satisfied. If one
inequality is not satisfied, then the point lies outside of the
region. We therefore describe a region as the difference between its
convex hull and a set of convex obstacles. As there are typically a
large number of obstacles with a large number of describing
inequalities, we simplify the description.  Here, it is most important
that the simplified representation does not include any additional
points. We simplify the representation by transforming the described
region into a slightly smaller sub-region that can be described with
considerably fewer inequalities (see
Section~\ref{sec:simplification}).

From the simplified representation of the feasible areas, we can now
deduce a linear program of manageable size and start enumerating the
packing patterns, as described in Section~\ref{sec:enumeration}. The
algorithm can be summarized as follows.

\begin{enumerate}
\item For each box in each orientation determine the region of
feasible points for the centers of the boxes.
\item For each region of feasible points $F$, compute its convex hull
  $C$ and a decomposition of $C-F$ into convex pieces.
\item Simplify the description.
\item Enumerate feasible packing patterns.
\end{enumerate}

Note that we restrict our experiments to the first step of the packing
routine, i.e., finding an optimal packing for boxes A--G, and even in
this step we ignore the golf bag. Ignoring the golf bag has no
technical reasons. We compare our results to the results given i
left out the golf bag in their experiments, the

\section{Computation of the Feasible Area}
\label{sec:feasiblearea}

\begin{figure}[t]
\center \includegraphics[width=0.95\textwidth]{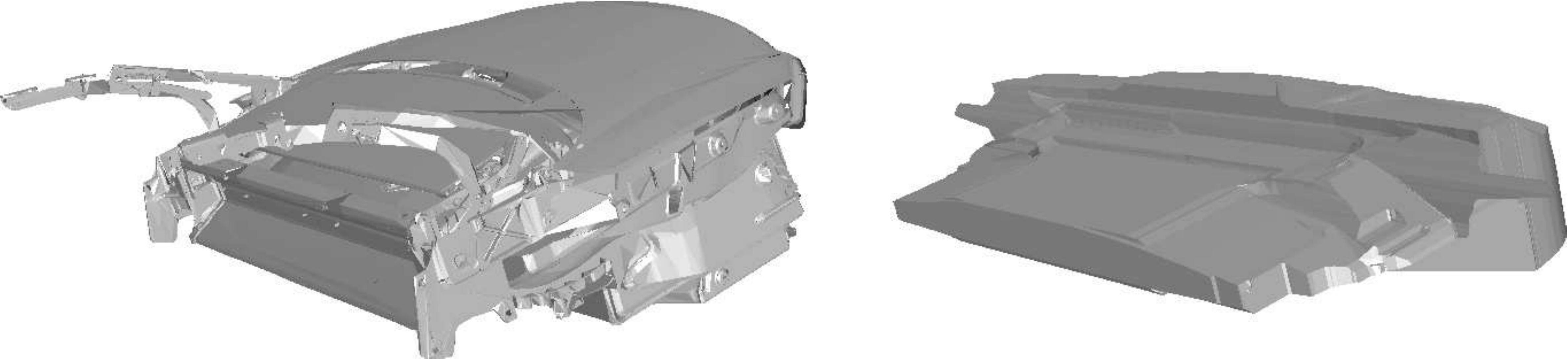}
\caption{
\label{fig:csp}
Trunk and the free space of box E (the orientation is fixed) with
respect to the trunk. Note that the right picture is enlarged to give
a better depiction of the free space.}
\end{figure}

A \emph{configuration space} of an object $P$ defines the set of all
placements of $P$. The placements may consist of a position and an
orientation in a $d$-dimensional work space, with respect to a set of
obstacles $Q$~\cite{bkos-cgaa-97}. The placement of $P$ is usually
given as the position of a reference point of $P$. This point can be
arbitrary.  In case of the trunk packing problem it is convenient to
use a corner of a suitcase, or, as we do it, the center of the
suitcase. The configuration space decomposes into the \emph{forbidden
space}, the set of configurations in which $P$ intersects with $Q$,
and the \emph{free space}, the set of configurations in which $P$ does
not intersect $Q$.

We restrict ourselves to a variant of the general problem, where the
orientation of the suitcases is limited to the six axis-aligned
directions. If the orientation of the object $P$ is fixed, e.g., $P$ is
a translational robot, the free space is a subset of the
three-dimensional space, which we denote as the \emph{feasible area}
of $P$. It can be computed as the complement of the Minkowski sum
$-P\oplus Q =\{p+q:p\in -P, q \in Q \}$, where $-P=\{-p\ |\ p\in P\}$ is
the inverted point set of $P$. 

In the trunk packing scenario $P$ is a suitcase and $Q$ is the
complement of the trunk's interior, i.e., the trunk's boundary and the
region outside the trunk. We compute the feasible area of a suitcase
with respect to a trunk as six complements of Minkowski sums---one for
each orientation. Figure~\ref{fig:csp} shows an example for the
Minkowski sum of a trunk and suitcase.

The Minkowski sum of two polyhedra is usually computed by the
so-called decomposition method. It decomposes the two polyhedra into
convex pieces, computes all pairwise Minkowski sums of the convex
pieces, and merges the pairwise sums. In the trunk-packing scenario,
one of the polyhedra, the complement of the trunk's interior, is an
infinite set. Unfortunately, none of the few existing implementations
of the decomposition method supports infinite point
sets~\cite{h-emspe-07,vm-amsap-04}. Also, the given input files
describe the trunks' boundary as a surface, which usually has
holes. Therefore the trunk cannot be modeled as a solid, which is a
necessary precondition for the decomposition method. Unfortunately, we
also do not know how to reliably remove the holes from a surface. It
seems to be a surprisingly complex problem.

\begin{figure}[t]
\center
\includegraphics[width=0.8\textwidth]{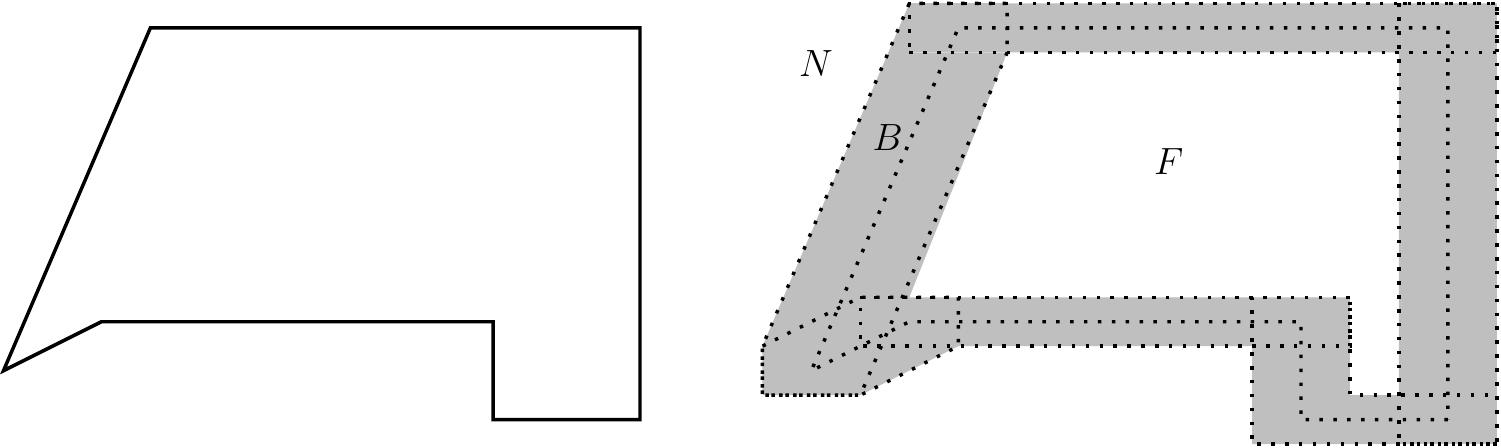}
\caption{\label{fig:surfaceMink} The Minkowski sum of the trunk on the
left and a box subdivides the space into three regions, as shown on
the right. Region $F$ equals the complement of the Minkowski sum of
the trunk's interior and the (inverted) box.}
\end{figure}

An alternative approach starts by computing the Minkowski sum of the
suitcase and the boundary of the trunk, which can be expressed as the
union of the Minkowski sums of the suitcase and a facet of the trunk's
boundary. If the holes in the surface are not too large the resulting
polyhedron $B$ decomposes the three-dimensional space into three
regions: $B$ itself, the region $N$ outside of $B$, and a void $F$
enclosed by $B$ (see Figure~\ref{fig:surfaceMink}). It is easy to see,
that $F$ coincides with the feasible area.

The trunk models are described by a set of triangles. We compute $B$
as the union of the Minkowski sums of the box and each triangle
defining the trunk. The Minkowski sum of a box and a triangle can
easily be computed as the convex hull of the vector sums of their
vertices.  For the union operation we use 3D Nef
polyhedra~\cite{hkm-bosnc-07} provided by {\cgal}~\cite{cgalhome}. The
use of {\cgal} is motivated by robustness issues caused by the
multitude of degenerate situations that usually occur during the union
step. We also need {\cgal} in the next step of our algorithm for
decomposing a polyhedra into convex pieces, a step that also includes
lots of degeneracy problems.

\section{Decomposition and Simplification of the Feasible Area}
\label{sec:simplification}

Our enumeration algorithm creates and solves linear programs. One part
of the linear programs describes the feasible region. In a linear
program it is preferable to work with convex regions, which can be
described as the intersection of linear inequalities. We therefore
describe a feasible region as its convex hull minus a set of convex
obstacles. The convex hull and the difference of the convex hull and
the feasible region are computed by {\cgal} functions. Then we
decompose the difference into convex pieces with the decomposition
method described in~\cite{h-emspe-07}.

The data sets generated by the decomposition are too large to find
good solutions in reasonable time. In our experiments, the description
of the feasible area according to our data format usually comprises
several hundred obstacles. The convex hull and a few of the obstacles
usually comprise a few hundred half-spaces. The rest of the obstacles
have around 6 to 40 facets. Our largest data set has 4904 obstacles
and is described by a total of 72554 half-spaces. We therefore need to
simplify the data sets, without changing the represented point set too
much. Most importantly, we do not want the simplified point set to
contain any points outside the feasible area.

We simplify in two steps. In a first step, we merge adjacent
obstacles. Because such a union is usually non-convex, we replace the
merged obstacles by their combined convex hull. The convex hulls of
two adjacent obstacles is larger than the union of the replaced
obstacles. Because of the structure of the decomposition---it is a
vertical decomposition---it is unlikely that the convex hull of two
adjacent obstacles overlaps a third obstacle, and surely it cannot
overlap the region outside the convex hull of the feasible area. It
usually overlaps a part of the feasible area. Since we want to shrink
the represented point set as little as possible, we need ways to measure
the size of regions. For this purpose we adapted the code of Mirtich's
volume integration method~\cite{m-facpm-96}, such that it computes the
volume of a cell of a Nef polyhedron. We replace two obstacles by
their combined convex hull, if the growth of the convex hull does not
exceed a given relative bound of $X\%$ or a given absolute bound of
$Ymm^3$. The replacement process is performed iteratively as long as
we find obstacle pairs that can be joined according to the given
percentage and the given fixed volume. Note, that we maintain the
volume of the union of each set of joined obstacles. Thus, to test for
replacement, we always compare the volume of a convex hull of a set of
obstacles, with the volume of their union. To find candidate obstacle
pairs, we just iterate in random order over all facets separating two
obstacles and check whether they qualify for replacement.

In a second step we reduce the number of describing facets of an
obstacle. We iteratively drop facets if as a result the size of the
obstacle does not increase much. The growth of an obstacle is measured
as the largest distance between a point in the enlarged obstacle and
the ignored facet. This quantity can be easily computed by using an
LP solver.

\section{Enumeration of Feasible Packings}
\label{sec:enumeration}

\begin{figure}[t]
\center
\includegraphics[width=0.8\textwidth, height=0.40\textwidth]{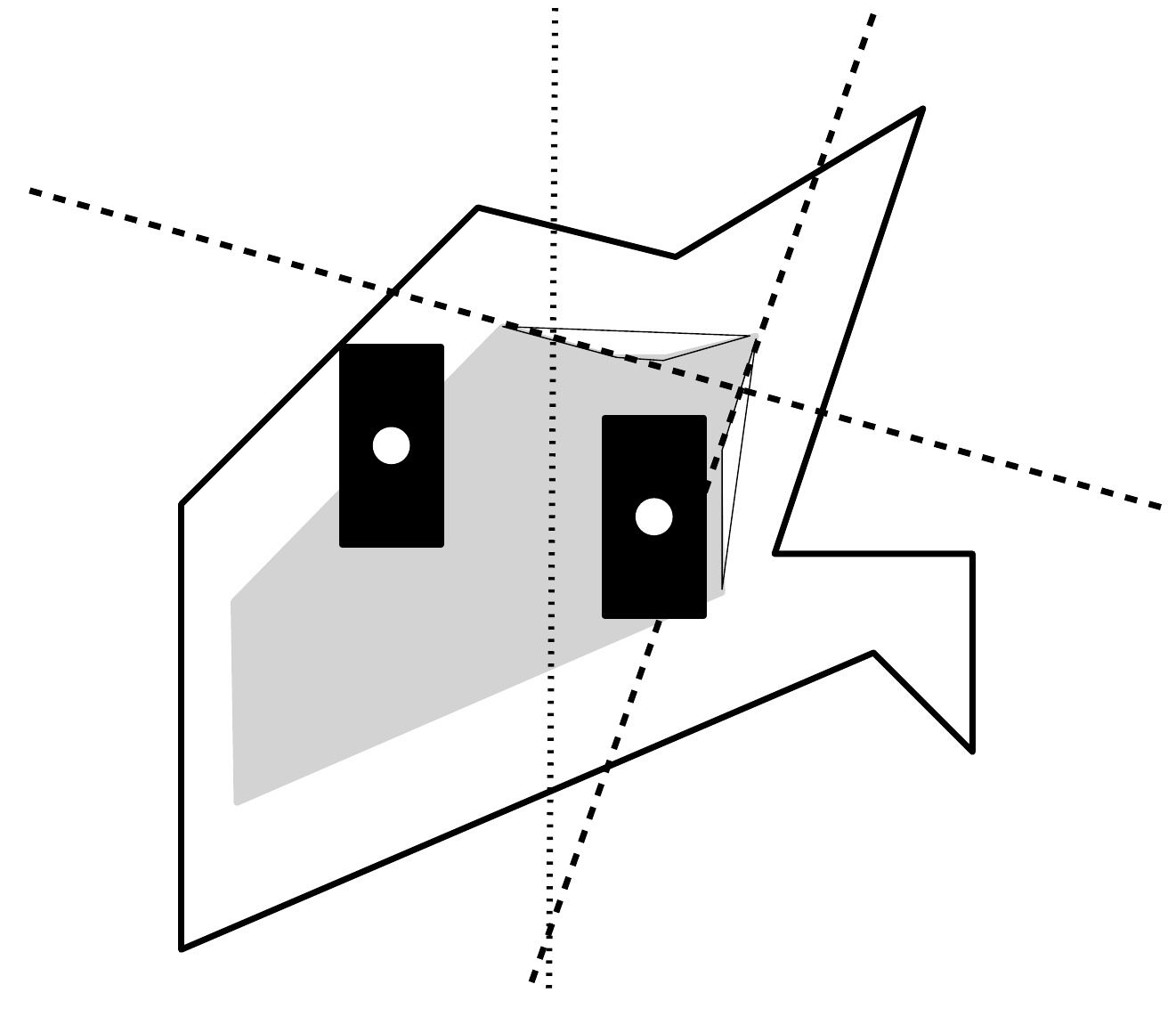}
\caption{
\label{fig:pattern}
  The feasible region of the black rectangle with respect to the
  (exterior of the) polygon is drawn in gray. It can be represented as
  the difference of its convex hull and two convex obstacles. The two
  depicted placements of the rectangle lie within the feasible region
  as their center points lie within the convex hull of the gray region
  and fail at least one facet inequality of each obstacle (dashed
  lines). The dotted line is a separating line of the two placements.}
\end{figure}

We find packings with the help of a linear-program solver (LP solver).
The linear programs that we specify consist of two sets of
constraints: constraints that define the convex hull of the
configuration spaces, and a packing pattern. A packing pattern
specifies a set of boxes together with their orientations, and one
separating plane for each box--box and each box--obstacle pair (see
Figure~\ref{fig:pattern}). A separating plane is a constraint that
separates two geometric objects in the 3-dimensional space. In case of
the box-box pairs the separating plane is axis-aligned and therefore
resembles either a left-of/right-of, in-front-of/behind, or
above/under relation. In case of a box--obstacle pair, the constraint
is a facet of the obstacle. The set of all packing patterns is
infinite, but we only consider patterns, which are either feasible
themselves, or which become feasible by removing one box. We call such
a packing pattern a \emph{candidate pattern}.

\begin{figure}
\begin{codebox}
\Procname{$\proc{PatternEnumeration}(PartialPattern\ P)$}
\li solve LP of P
\li \If (LP is feasible) \Indentmore
\li \If (no intersections exist) \Indentmore
\li save packing if it is the largest found so far 
\li \For all boxes $b$ and all orientations $d$ \Indentmore
\li $\proc{PatternEnumeration}(P\cup b_d)$ \End
\li \Else \If(there is a box--box intersection)
\li  find box pair $(b_0, b_1)$ with largest intersection
\li \For all relative orders $r$ \Indentmore
\li compute constraint $c_r$ enforcing $r$ on $(b_0, b_1)$
\li $\proc{PatternEnumeration}(P\cup c_r)$ \End
\li \Else find box--obstacle pair $(b, o)$ with largest intersection
\li \For all defining facets of $o$ \Indentmore
\li compute constraint $c_o$ separating $b$ and $o$
\li $\proc{PatternEnumeration}(P\cup c_o)$ \End \End \End
\end{codebox}
\caption{\label{fig:patternEnumeration} Enumeration of partial packing
  pattern. The pattern in the first call specifies a box and its
  orientation. Constraints are only added in the recursive calls.}
\end{figure}

The set of candidate patterns is finite, but it is still too large to
find good packings in a reasonable time. We omit enumerating all
patterns as follows. We do not specify all constraints of a pattern,
but check subsets of pattern, further on denoted as \emph{partial
packing pattern}. If the LP solver cannot solve the linear program of
a partial pattern, we already know that the partial pattern cannot be
extended into a solvable pattern. If the LP solver provides a solution
to the partial pattern, this solution may not be a solution for the
complete pattern. There can still be intersections of box--box or
box--obstacle pairs, for which no constraint is specified, yet. We
identify those intersection, add a constraint to the partial pattern
that prevents one of them, and then let an LP solver solve this new
partial pattern. If the LP solver returns a solution without
intersections, we have found a packing and can try for a larger set of
boxes.

The order, in which we add constraints to a partial pattern, is
crucial in keeping the search space as small as possible. It is easy
to argue that it is much more efficient to first resolve all box--box
intersections and then go on with the box--obstacle intersections than
proceeding vice versa. As soon as relations between most of the boxes
are specified, the boxes form a big rigid object. Going on with
constraints for the box--obstacle pairs moves around this big object
until it fits, or until the LP solver cannot provide a solution any
more. Most of the times, we expect only few movements until a partial
pattern becomes unsolvable. On the other hand, if we start with
resolving box--obstacle intersections, the search tree will always be
huge. Without constraints for the box--box relations, the
box--obstacle constraints will only cause the boxes to overlap one
another. We cannot expect to find an unsolvable partial pattern
without specifying the box--box relations.

The algorithm is summarized in Figure~\ref{fig:patternEnumeration}. We
left out the details that are necessary to omit redundant
enumerations, and the branch-and-bound techniques that omit testing
uninteresting packings.

\ignore{
In this fashion we add the boxes one by one starting with the
largest. If we were able to make the partial packing pattern feasible,
we try to add another box. Thereby, we compute an upper bound on the
best possible packing that can emerge from the partial packing pattern
by assuming that the maximum number of remaining smaller boxes can be
packed. If this upper bound is smaller than the best known packing, we
stop the search in this branch. The computation of better upper bounds
would be a major step in improving the algorithm.
}

\section{Experiments}
\label{sec:experiments}

We implemented our approach in {\CC} and performed the computations on
a machine with a 2.4 GHz AMD Opteron 250 processor and 4 GB RAM. All
linear programs were solved with CPLEX~10.0~\cite{CPLEX}. We tested
our algorithm on three different trunk models, {\TSmall} (99352
triangles), {\TMedium} (66743 triangles), and {\TLarge} (42182
triangles), which are the same that were used in previous
experiments~\cite{bsw,efkrs-05}.

\ignore{
\begin{table}
\begin{center}
\begin{tabular}{|l|r|r|r|}
\hline
trunk & triangles & manual packing & 
\hline
\TSmall & 99352 & $5.34 \ft^3$ & 
\TMedium & 66743 & $9.36 \ft^3$ & 
\TLarge & 42182 & $11.84 \ft^3$ & 
\hline
\end{tabular}
\vspace{3mm}
\caption{The volume of the best packings found without using
the H-boxes by manual packing, by the grid-based method using physics
simulation, and by our approach.}
\label{table:trunkData}
\end{center}
}

\begin{table}[b]
\small
\center
\begin{tabular}{|c|cc|cccccc|cccc|cccc|}
\hline
box
& \multicolumn{2}{|c|}{A}
& \multicolumn{6}{|c|}{B}
& \multicolumn{4}{|c|}{C}
& \multicolumn{4}{|c|}{D} \\
orientation & yxz & xyz & 
zyx & zxy & yzx & xzy & yxz & xyz &
yzx & xzy & yxz & xyz &
yzx & xzy & yxz & xyz \\
\hline
volume $[\dm^3]$ & 21.7 & 17.3 & 2.5 & 1.3 & 40.1 & 30.0 & 74.2 & 67.0 & 6.4 & 1.4 & 24.3 & 15.5 & 1.0 & 0.2 & 33.0 & 29.6 \\
facets $[10^3]$& 11.4 & 11.8 & 4.5 & 3.2 & 12.0 & 10.9 & 23.3 & 24.0 & 7.3 & 1.7 & 10.2 & 11.2 & 1.9 & 0.7 & 14.1 & 14.5 \\
\hline
\end{tabular}
\vspace{3mm}
\begin{tabular}{|c|cccccc|cccc|cccccc|}
\hline
& \multicolumn{6}{|c|}{E}
& \multicolumn{4}{|c|}{F}
& \multicolumn{6}{|c|}{H} \\
& zyx & zxy & yzx & xzy & yxz & xyz &
yzx & xzy & yxz & xyz &
zyx & zxy & yzx & xzy & yxz & xyz \\
\hline
volume $[\dm^3]$ & 31.1 & 30.3 & 84.3 & 74.9 & 90.7 & 82.1 & 25.6 & 15.4 & 55.3 & 46.0 & 68.2 & 69.4 & 138.4 & 127.0 & 130.6 & 118.1 \\
facets $[10^3]$ & 12.5 & 12.7 & 20.4 & 17.8 & 24.7 & 22.1 & 10.7 & 8.9 & 19.8 & 20.4 & 17.2 & 18.2 & 35.6 & 31.5 & 32.4 & 28.7 \\
\hline
\end{tabular}
\vspace{0.5mm}
\caption{\label{tab:feasibleAreas} 
Size of the non-empty feasible areas of {\TMedium} and complexity of
their description. }
\end{table}

The computation of the feasible area of a single suitcase (for one
orientation) takes 66 minutes for trunk {\TSmall}, 36 minutes for
{\TMedium}, and 69 minutes for trunk {\TLarge}. The running times do
not vary much for different suitcases or orientations. Even the
determination of an empty feasible area needs the full computation
time. Table~\ref{tab:feasibleAreas} lists the volumes and the number
of describing facets for the non-empty feasible areas of all boxes in
all six orientations with respect to trunk {\TMedium}. In the tables
we denote the six orientations by permutations of the three coordinate
axis. The first letter of a permutation represents the alignment of
the longest box side, the third letter represents the alignment of
the shortest box side.

For the decomposition and the simplification we did not explicitly
measure the running times. From the dates of the created files we can
see that decomposition and simplification of a single feasible area
takes 1--5 minutes; a whole set of feasible areas of all boxes and all
orientations is decomposed and simplified within 2.5--3.5 hours.

\begin{table}[t]
\center
\begin{tabular}{|cc|ccccc|ccccc|ccccc|}
\hline
& & \multicolumn{5}{|c|}{$1000 mm^3$} & \multicolumn{5}{|c|}{$10000 mm^3$} & \multicolumn{5}{|c|}{$100000 mm^3$} \\
\raisebox{1.4ex}[0cm][0cm]{box} &
\raisebox{1.4ex}[0cm][0cm]{orient.} & 
$5\%$ & $10\%$ & $20\%$ & $50\%$ & $80\%$ & $5\%$ & $10\%$ & $20\%$ & $50\%$ & $80\%$
& $5\%$ & $10\%$ & $20\%$ & $50\%$ & $80\%$ \\
\hline
A & yxz & 99.5 & 98.8 & 97.9 & 94.4 & 85.1 & 98.8 & 98.3 & 97.5 & 94.1 & 86.0 & 93.8 & 93.8 & 93.8 & 91.6 & 84.0 \\
A & xyz & 99.7 & 99.1 & 97.9 & 92.5 & 86.6 & 98.8 & 98.4 & 97.6 & 92.8 & 86.5 & 91.9 & 91.9 & 91.9 & 89.6 & 84.6 \\
H & zyx & 99.8 & 99.5 & 98.6 & 94.6 & 91.8 & 99.5 & 99.4 & 98.9 & 94.6 & 91.9 & 97.1 & 97.1 & 97.1 & 94.7 & 91.7 \\
H & zxy & 99.8 & 99.5 & 98.6 & 95.4 & 92.6 & 99.5 & 99.3 & 98.5 & 95.2 & 92.9 & 96.8 & 96.8 & 96.6 & 94.5 & 91.1 \\
H & yzx & 99.8 & 99.4 & 98.4 & 94.6 & 90.8 & 99.6 & 99.3 & 98.3 & 94.9 & 90.8 & 97.7 & 97.7 & 97.4 & 93.7 & 90.2 \\
H & xzy & 99.8 & 99.6 & 98.8 & 95.6 & 93.2 & 99.5 & 99.3 & 98.7 & 95.5 & 93.5 & 97.8 & 97.8 & 97.7 & 95.2 & 93.1 \\
H & yxz & 99.7 & 99.2 & 98.1 & 95.0 & 90.5 & 99.5 & 99.1 & 97.9 & 95.1 & 90.4 & 97.6 & 97.5 & 97.2 & 94.5 & 88.9 \\
H & xyz & 99.8 & 99.6 & 98.9 & 96.0 & 93.2 & 99.5 & 99.4 & 98.7 & 96.4 & 93.1 & 97.9 & 97.9 & 97.6 & 95.9 & 93.0 \\
\hline
\multicolumn{2}{|c|}{average} 
& 99.7 & 99.4 & 98.5 & 94.9 & 91.3 
& 99.4 & 99.1 & 98.3 & 95.0 & 91.3 
& 96.6 & 96.6 & 96.3 & 93.9 & 90.2 \\
\hline
\end{tabular}
\vspace{3mm}
\begin{tabular}{|cc|ccccc|ccccc|ccccc|}
\hline
& & \multicolumn{5}{|c|}{$1000 mm^3$} & \multicolumn{5}{|c|}{$10000 mm^3$} & \multicolumn{5}{|c|}{$100000 mm^3$} \\
\raisebox{1.4ex}[0cm][0cm]{box} &
\raisebox{1.4ex}[0cm][0cm]{orient.} & 
$5\%$ & $10\%$ & $20\%$ & $50\%$ & $80\%$ & $5\%$ & $10\%$ & $20\%$ & $50\%$ & $80\%$
& $5\%$ & $10\%$ & $20\%$ & $50\%$ & $80\%$ \\
\hline
A & yxz & 27.2 & 23.1 & 19.1 & 14.5 & 12.3 & 22.9 & 20.1 & 18.2 & 15.0 & 13.5 & 15.1 & 15.1 & 14.8 & 13.3 & 11.6 \\
A & xyz & 25.5 & 23.5 & 18.8 & 14.4 & 11.9 & 19.4 & 18.8 & 17.5 & 14.8 & 12.7 & 13.5 & 13.5 & 13.5 & 13.3 & 11.6 \\
H & zyx & 26.1 & 22.2 & 17.0 & 10.7 & 8.3 & 22.5 & 21.5 & 18.2 & 10.7 & 8.3 & 14.7 & 14.7 & 14.1 & 11.7 & 9.1 \\
H & zxy & 26.9 & 21.1 & 17.0 & 11.4 & 8.5 & 21.7 & 19.5 & 17.2 & 11.7 & 8.5 & 14.4 & 14.3 & 13.7 & 11.7 & 9.0 \\
H & yzx & 24.8 & 19.6 & 14.6 & 9.3 & 7.5 & 20.7 & 17.9 & 14.2 & 10.1 & 7.4 & 13.8 & 13.5 & 12.7 & 9.3 & 7.9 \\
H & xzy & 23.2 & 19.3 & 14.4 & 9.2 & 8.4 & 19.6 & 17.8 & 14.7 & 10.0 & 8.6 & 12.5 & 12.3 & 11.3 & 8.9 & 8.6 \\
H & yxz & 24.5 & 18.4 & 14.3 & 9.8 & 8.5 & 20.3 & 16.9 & 14.1 & 9.6 & 8.6 & 14.1 & 14.0 & 13.5 & 10.6 & 8.2 \\
H & xyz & 23.7 & 19.7 & 15.6 & 10.8 & 8.7 & 19.4 & 17.9 & 14.7 & 11.0 & 9.3 & 12.4 & 12.4 & 11.4 & 10.0 & 8.8 \\
\hline
\multicolumn{2}{|c|}{average} 
& 25.4 & 21.2 & 17.1 & 11.5 &  9.8
& 21.0 & 18.9 & 16.5 & 11.7 &  9.8
& 14.0 & 13.8 & 13.2 & 11.2 &  9.8 \\
\hline
\end{tabular}

\caption{Impact of first simplification step on the non-empty feasible
  areas of A and H-box with respect to trunk {\TMedium} using
  parameter values $X=5\%,10\%, 20\%, 50\%, 80\%$ (maximum percentual
  growth of obstacle union) and $Y=1000 mm^3, 10000 mm^3, 100000 mm^3$
  (maximum absolute growth of obstacle union). Top: Volume of the
  represented area after the simplification relative to original
  volume.  Bottom: Number of facets used for describing the feasible
  area after the simplification relative to original number of
  facets.}

\label{table:simplificationDetailed}
\end{table}

Table~\ref{table:simplificationDetailed} shows the effectiveness of
our first simplification step. Applying it to trunk {\TMedium}, we can
reduce the number of defining half-spaces to around 20\% while we only
loose 1--2\% of the feasible area. This trade-off is a bit better for
{\TLarge} and a bit worse for {\TSmall}.

\begin{table}[t]
\center
\begin{tabular} {|c|cccccccc|}
\hline
allowed growth $[mm]$ & 0 & 0.1 & 0.2 & 0.5 & 1 & 2 & 5 & 10 \\ 
obstacle facets & $100\%$ & $58\%$ & $51\%$ & $41\%$ & $33\%$ & $27\%$ & $21\%$ & $18\%$ \\
\hline
\end{tabular}
\vspace{1mm}
\caption{ \label{table:secondSimplificationStep} Effect of second
simplification step applied to the (simplified) feasible area of trunk
{\TSmall} and suitcase D with orientation $zxy$.}
\end{table}

Table~\ref{table:secondSimplificationStep} shows the influence of the
second simplification step. On average it additionally reduces the
number of defining half-spaces by about $40\%$, $65\%$, or $80\%$, if
we allow a growth by $0.1mm$, $1mm$ or $10mm$, respectively. 

The enumeration phase of our algorithm does not terminate within $24$
hours for the tested trunks. We terminate the enumeration step after
$24$ hours and report the volume of the best
packing. Table~\ref{table:quality} compares our results with the
results of previous approaches. For the trunk {\TSmall}, we found a
better packing than anyone else. For the two other trunks we are not
as good as the other approaches. It seems that we perform better when
the trunk volume is comparably small. This is probably due the limited
number of feasible partial packing patterns that we can enumerate in
the given time. On the other hand, the best packing was usually found
within one hour. Up to now, we enumerate the packing patterns in a
depth-first manner. Changing the order in a clever way, e.g., by
determining a measure how promising a certain partial packing pattern
is, we expect that the performance of our approach can be improved.

\begin{table}[t]
\begin{center}
\begin{tabular}{|l|r|r|r|}
\hline
trunk & manual & physics sim. & Minkowski sum \\
\hline
\TSmall  & $151.1 \dm^3$ & $159.9 \dm^3$ & $163.9 \dm^3$\\
\TMedium & $264.9 \dm^3$ & $266.6 \dm^3$ & $263.5 \dm^3$\\
\TLarge  & $335.1 \dm^3$ & $358.6 \dm^3$ & $321.8 \dm^3$\\
\hline
\end{tabular}
\vspace{3mm}
\caption{The volume of the best packings found without using
the H-boxes by manual packing, by the grid-based method using physics
simulation, and by our approach.}
\label{table:quality}
\end{center}
\end{table}

Summarizing the running times of the four steps of our algorithm, the
computation of the feasible area takes 35--70 minutes for a single box
with fixed orientation, which sums up to a total of 24.5--49 hours for
all 42 combinations. Then the decomposition and simplification of a
complete set takes 2.5--3.5 hours. Finally, the enumeration will
probably yield a good solution within 1 hour. In total the computation
can last more than 50 hours. The running times become more interesting
if we perform computations in parallel. Having six machines instead of
one, we can compute the feasible areas in less than 8.5 hours, then
create six different simplified point sets using different parameters,
and can expect to have a good solution within 13 hours.

\section{Conclusion}
\label{sec:conclusion}

We have given a new algorithm to compute the volume of a trunk
according to the SAE J110 standard as used in the USA. It outperforms
previous approaches for small trunks.

In spite of the improvable performance of the computation of the
feasible areas, our approach already meets the demands of the German
car manufacturer, since many tasks can be performed in parallel. With
six machines like ours it is possible to compute the feasible areas
and a few sets of their simplified representations on the first half
of a day. The other half day suffices to give the LP solver the
opportunity to come up with good packings.

We outlined several lines of further research to improve the
performance of our algorithm. We are interested in reliable methods to
close the surface of a trunk, such that we can design heuristics to
speed up the computation of the feasible regions. Furthermore we want
to improve the enumeration phase by the development of better upper
bounds for the volume that can be obtained by extending a given
partial packing pattern together with algorithms to compute these
upper bounds and by clever heuristics to scan the large search space.

\bibliographystyle{plain}
\bibliography{trunk}

\begin{thebibliography}{10}

\bibitem{bsw}
E.~Althaus, T.~Baumann, E.~Sch{\"o}mer, and K.~Werth.
\newblock Solving geometric packing problems based on physics simulation.
\newblock Unpublished.

\bibitem{absw-07}
E.~Althaus, T.~Baumann, E.~Sch{\"o}mer, and K.~Werth.
\newblock Trunk packing revisited.
\newblock In {\em Proceedings of the 6th International Workshop on Experimental
  and Efficient Algorithms, (WEA 2007)}, Lecture Notes in Computer Science,
  Rome, Italy, 2007.

\bibitem{cd-03}
J.~Cagan and Ding Q.
\newblock Automated trunk packing with extended pattern search.
\newblock In {\em Virtual Engineering, Simulation {\&} Optimization}, 2003.

\bibitem{cgalhome}
{\em The CGAL Homepage}.
\newblock \path|http://www.cgal.org/|.

\bibitem{CPLEX}
{ILOG CPLEX 10.0}.
\newblock \path|http://www.ilog.com/|.

\bibitem{bkos-cgaa-97}
M.~de~Berg, M.~van Kreveld, M.~Overmars, and O.~Schwarzkopf.
\newblock {\em Computational Geometry: Algorithms and Applications}.
\newblock Springer Verlag, 1997.

\bibitem{efkrs-05}
F.~Eisenbrand, S.~Funke, A.~Karrenbauer, J.~Reichel, and E.~Sch\"{o}mer.
\newblock Packing a trunk: now with a twist!
\newblock In {\em SPM '05: Proceedings of the 2005 ACM Symposium on Solid and
  Physical Modeling}, pages 197--206, New York, NY, USA, 2005. ACM.

\bibitem{Reichel2003}
F.~Eisenbrand, S.~Funke, J.~Reichel, and E.~Sch{\"o}mer.
\newblock Packing a trunk.
\newblock In Giuseppe Di~Battista and Uri Zwick, editors, {\em Algorithms - ESA
  2003: 11th Annual European Symposium}, volume 2832 of {\em Lecture Notes in
  Computer Science}, pages 618--629, Budapest, Hungary, September 2003.
  Springer.

\bibitem{h-emspe-07}
P.~Hachenberger.
\newblock Exact {Minkowksi} sums of polyhedra and exact and efficient
  decomposition of polyhedra in convex pieces.
\newblock In {\em 15th European Symposium on Algorithms (ESA'07)}, pages
  669--680, 2007.

\bibitem{hkm-bosnc-07}
P.~Hachenberger, L.~Kettner, and K.~Mehlhorn.
\newblock Boolean operations on {3D} selective {Nef} complexes: Data structure,
  algorithms, optimized implementation and experiments.
\newblock {\em Computational Geometry: Theory and Applications},
  38(1--2):64--99, 2007.

\bibitem{Karr04}
A.~Karrenbauer.
\newblock Packing boxes with arbitrary rotations.
\newblock Master's thesis, Universit{\"a}t des Saarlandes, 2004.

\bibitem{m-facpm-96}
B.~Mirtich.
\newblock Fast and accurate computation of polyhedral mass properties.
\newblock {\em Journal of Graphics Tools}, 1(2):31--50, 1996.

\bibitem{Neum06}
U.~Neumann.
\newblock Optimierungsverfahren zur normgerechten {Volumenbestimmung} von
  {Kofferr\"aumen} im europ\"aischen {Automobilbau}.
\newblock Master's thesis, Technische Universit{\"a}t Braunschweig, 2006.

\bibitem{Reichel06}
J.~Reichel.
\newblock {\em Combinatorial Approaches for the Trunk Packing Problem}.
\newblock PhD thesis, Universit{\"a}t des Saarlandes, July 2006.

\bibitem{Ries05}
J.~Rieskamp.
\newblock Automation and optimization of monte carlo based trunk packing.
\newblock Master's thesis, Universit{\"a}t des Saarlandes, 2005.

\bibitem{schepers}
J.~Schepers.
\newblock {\em Exakte Algorithmen f¨ur orthogonale Packungsprobleme}.
\newblock PhD thesis, K{\"o}ln, 1997.

\bibitem{vm-amsap-04}
G.~Varadhan and D.~Manocha.
\newblock Accurate {Minkowski} sum approximation of polyhedral models.
\newblock In {\em Proc. Comp. Graphics and Appl., 12th Pacific Conf. on
  (PG'04)}, pages 392--401. IEEE Computer Society, 2004.

\end{thebibliography}
\end{document}